\documentclass[]{elsarticle}

\usepackage{graphicx}
\usepackage{epsfig}
\usepackage{amssymb}

\begin{document}

\begin{frontmatter}

\title{Test in a beam of large-area Micromegas chambers for sampling calorimetry}

\author[A]{C.~Adloff}
\author[A]{M.~Chefdeville\corref{cor1}}\cortext[cor1]{Corresponding author}
\ead{chefdevi@lapp.in2p3.fr}
\author[A]{A.~Dalmaz}
\author[A]{C.~Drancourt}
\author[A]{R.~Gaglione}
\author[A]{N.~Geffroy}
\author[A]{J.~Jacquemier}
\author[A]{Y.~Karyotakis}
\author[A]{I.~Koletsou}
\author[A]{F.~Peltier}
\author[A]{J.~Samarati}
\author[A]{G.~Vouters}

\address[A]{LAPP, Laboratoire d'Annecy-le-Vieux de Physique des Particules, Universit\'{e} de Savoie, CNRS/IN2P3, 9 Chemin de Bellevue - BP 110, 74941 Annecy-le-Vieux Cedex, France}

\begin{abstract}
Application of Micromegas for sampling calorimetry puts specific constraints on the design and performance of this gaseous detector. In particular, uniform and linear response, low noise and stability against high ionisation density deposits are prerequisites to achieving good energy resolution. A Micromegas-based hadronic calorimeter was proposed for an application at a future linear collider experiment and three technologically advanced prototypes of 1$\times$1\,m$^{2}$ were constructed. Their merits relative to the above-mentioned criteria are discussed on the basis of measurements performed at the CERN SPS test-beam facility.
\end{abstract}

\end{frontmatter}

\section{Introduction}

\subsection{Micromegas for sampling calorimetry}

The linear response of Micromegas is an advantage for sampling calorimetry. This property originates from a fast collection of avalanche ions at the mesh (25--100\,ns), a relatively low operating gas gain (a few thousands) and a small transverse diffusion of avalanche electrons ($\sim$\,10--20\,$\mu$m RMS) \cite{GIO96,COL04}. In this work, an application for Particle-Flow hadron calorimetry at a future linear collider is targeted (ILC or CLIC \cite{ILC,CLIC}). The foreseen calorimeter is highly segmented and called the Semi-Digital Hadron CALorimeter (or SDHCAL). Its expected performance can only be achieved with stringent constrains on the mechanics and electronics of the sampling layers \cite{TDR}. Three Micromegas prototypes fulfulling some of the constraints were built \cite{CHE13}. Their main features are a large sensitive area ($\sim$\,1$\times$1\,m$^{2}$), a compact mechanical design (Bulk Micromegas \cite{BULK}, embedded ASICs) and ILC-specific front-end electronics (self-triggering, power-pulsing).

The energy resolution of sampling calorimeters is limited by intrinsic and sampling fluctuations \cite{WIG00}. Depending on the energy range of interest, noise and non-uniformity of the active layer characteristics may also impact on the resolution \cite{BUD95}. In the case of a gaseous medium, calorimeter response and linearity can suffer from particle rates as a result of a competition between applied and space-charge fields \cite{ATA83}. Moreover, nuclear fragments produced in hadron showers may lead to very large signals \cite{GAL86} and potentially discharges, spoiling the energy measurement. These effects were studied in testbeams using 3 large-area Micromegas prototypes. When possible, the results are extrapolated to the case of a Micromegas-based calorimeter.

\section{Experimental setup}

Several measurements are presented in this paper. They were performed at CERN in November 2012 with two different experimental setups installed in the beam lines H4 and H2 of the SPS north hall. The setup in H4 is composed of 3 Micromegas prototypes of 1$\times$1\,m$^{2}$ and an absorber, it will be referred to as the standalone setup. The second one installed in H2 consists of the CALICE Fe-SDHCAL prototype \cite{PUE11} equipped with 46 RPCs and 3 Micromegas layers. After a description of the Micromegas prototypes and setups, uniformity measurements performed with muons are presented. They are followed by noise and stability studies in muon and pion beams. Each time, the use of a given setup is justified. During the tests, the Micromegas prototypes were flushed with a gas mixture of Ar/CF$_{4}$/\textit{i}C$_{4}$H$_{10}$ 95/3/2 at a few mbar over atmospheric pressure. The drift field is kept at 300\,V/cm at which the electron drift velocity in the gas mixture reaches a local maximum ($\sim$\,8\,cm/$\mu$s).

\subsection{The 1$\times$1\,m$^{2}$ Micromegas prototype}

\begin{figure}[htbp]
\centering
\includegraphics[width=0.8\columnwidth]{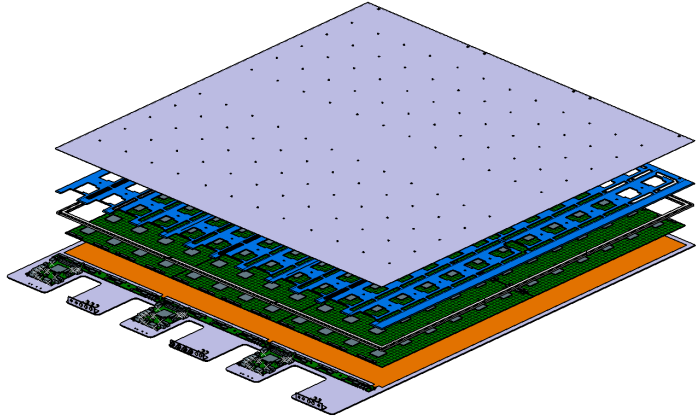}
\caption{Drawing of the 1$\times$1\,m$^{2}$ prototype showing from bottom to top: a steel baseplate with 3 patch-panels and readout boards (DIF and inter-DIF), a drift electrode glued on the baseplate, 6 Active Sensor Units each equipped with 24 ASICs, a mask and a removable cover.}
\label{m2proto_design}
\end{figure}

A detailed description of the prototypes can be found in \cite{CHE13}. Only a few points essential to the understanding of the reported measurements are recalled here.  The 1$\times$1\,m$^{2}$ Micromegas prototype consists of 6 printed circuit boards, each one equipped with a Bulk mesh, 1536 anode pads of 1$\times$1\,cm$^{2}$ arranged in a 32$\times$48 matrix and 24 MICROROC ASICs. The 6 boards called Active Sensor Units (or ASU) are placed inside a single gas chamber (Fig.\,\ref{m2proto_design}). The ASICs perform the amplification of the pad signals, shaping and discrimination (2-bit encoding or 3 thresholds). The result of the discrimination is recorded in the ASIC memory with a timestamp given by a 5\,MHz clock. The memory content is read out when the ASICs receive a trigger signal from the data acquisition system. This signal is generated externally by \textit{e.g.} a triggering device or internally when one ASIC memory is full (the memory is 127 event deep). For detector characterisation purposes, a slow analogue readout of the shaper outputs is also possible. In this case, the digitisation is performed outside the detector by readout boards called DIF.

\subsection{Standalone setup}

\begin{figure}[htbp]
\centering
\includegraphics[width=0.8\columnwidth]{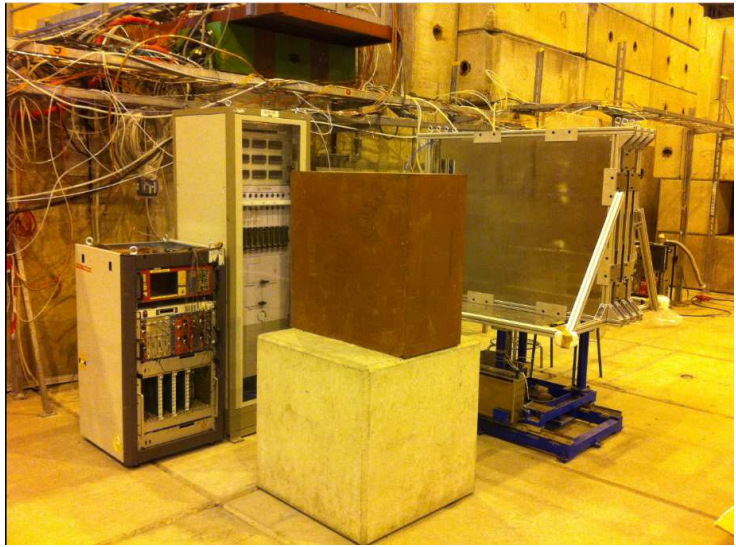}
\caption{Photograph of the standalone setup installed in line H4 of the SPS North Hall at CERN in November 2012. Following the beam direction, are shown from left to right: 2 racks for power supply, trigger electronics and gas distribution, an iron absorber block and a mechanical structure (mounted on a movable stage) that holds the Micromegas prototypes.}
\label{m2proto_h4line}
\end{figure}

The standalone setup allows to efficiently record many short runs. It is thus well suited for studying the effects of various experimental settings (\textit{e.g.} detector voltage, beam rate) on the response. Installed in the H4 line, it consists of a triggering device, an absorber and 3 Micromegas prototypes placed along the beam direction (Fig.\,\ref{m2proto_h4line}). The triggering device is composed of two scintillators and photomultiplier tubes held in mechanical structure which is fixed to the absorber. The absorber is a block of iron, 80$\times$80\,cm$^{2}$ area perpendicular to the beam direction and 40\,cm width ($\sim$\,2 interaction length $\lambda_{\rm int}$). Pions of 150\,GeV crossing the absorber in this direction will on average shower in its centre. At this energy, the hadronic shower maximum (also an average quantity) is reached about 1\,$\lambda_{\rm int}$ after the shower start, at the rear side of the block \cite{BLA09}. Therefore, this experimental setup is also used to study the detector stability to high-density energy deposits typical of hadronic showers.

\subsection{The CALICE SDHCAL}

The SDHCAL is a stack of 51 steel absorber plates of $\sim$\,1$\times$1\,m$^{2}$ and 1.5\,cm thickness separated by 1.3\,cm air gaps. During the tests in line H2, the SDHCAL was instrumented with 46 RPCs and 3 Micromegas of similar design and readout electronics. It was used to record hadron showers and muons. Due to the modest rate capability of the RPCs, this setup is too slow for performing voltage or rate scans. On the other hand, its large size and high granularity ($\sim$\,4.5$\times$10$^{5}$ pads of $\sim$\,1$\times$1\,cm$^{2}$) allow a precise position scan over the whole area of the Micromegas prototypes. Also, the high pattern recognition capability of the SDHCAL allows to distinguish hits due to noise from hits due to the interaction of particles in the calorimeter. Therefore, it is used for measuring noise rates. Along the direction of the beam, the Micromegas prototypes were inserted behind the absorbers number 20, 35 and 50.

\section{Uniformity of the muon response}

The response of 3 prototypes to high-energy muons was measured for different values of gas gain (or mesh voltage) and threshold, and in different regions of the prototype area. The gas gain dependence is first measured in a small region using the standalone setup. Then, the gain is fixed and the response is measured over the whole area inside the SDHCAL. Thanks to the 2-bit readout, the response is in fact determined at each run for 3 values of threshold.

\subsection{Voltage scan}

The scan is performed in several short runs in a 1\,kHz beam of 150\,GeV muons collimated to the size of the region read out by an ASIC (8$\times$8\,cm$^{2}$). At this energy, muons are a rough approximation of minimum ionising particles (MIPs), their energy loss being about 50\% larger \cite{PDG}. Multiple scattering, however, is minimal which is desirable for tracking studies. A first scan is performed with the analogue and digital readout to measure the Landau distribution. Then, efficiency and hit multiplicity are measured using the digital readout only.

\subsubsection{Analogue response}

\begin{figure}[htbp]
\centering
\includegraphics[width=0.75\columnwidth]{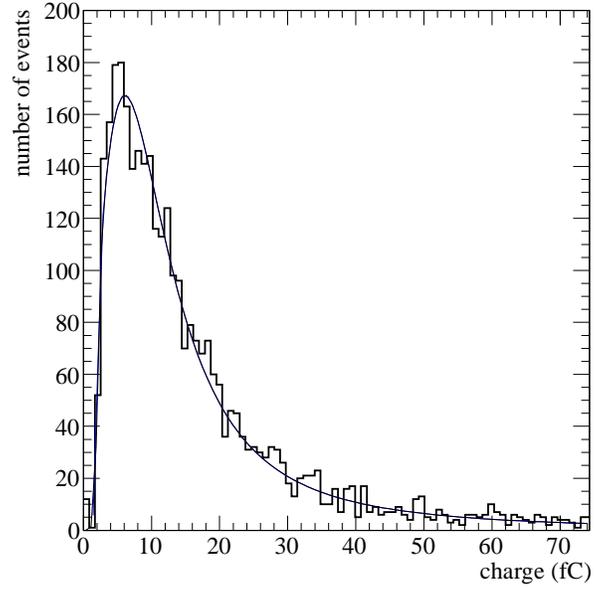}
\includegraphics[width=0.75\columnwidth]{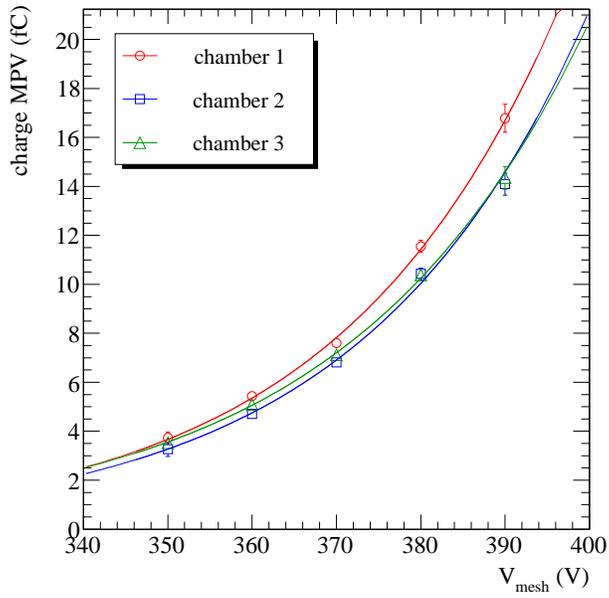}
\caption{Landau distribution from 150\,GeV muons at 370\,V (top). Most probable charge versus mesh voltage (bottom). Lines are the result of a fit where fit functions are the product of a Landau times as sigmoid (top) and an exponential (bottom).}
\label{voltage}
\end{figure}

The trigger signal formed by the time coincidence of the PMT signals is delayed to match the peaking time of the MICROROC ASIC shapers (200\,ns). When received at the DIFs, it is asynchronously forwarded to the ASICs. The shaper output voltages are held and subsequently read out and digitised by the DIF. Hits recorded in the ASIC memory are also read out. The mesh voltage was varied between 350--390\,V. At each voltage, a sample of $\sim$\,5$\times$10$^{3}$ muons was recorded.
After pedestal subtraction, only ADC values of the channels with a hit at the time of the trigger are summed. The sum generally involves only one hit and is converted into charge. The charge distribution measured in a prototype at 370\,V is shown in Fig.\,\ref{voltage} (top).

The most probable value (MPV) is determined using a fit and plotted versus mesh voltage for the 3 prototypes in Fig.\,\ref{voltage} (bottom). It follows the variations of the gas gain and is thus well described by an exponential function. At 370\,V, a most probable value of about 7\,fC is found for the 3 prototypes. The slope of the exponential function lies between 0.035--0.038\,/\,V, as expected in this gas mixture. It is similar for the 3 prototypes as well. In conclusion, the prototypes have a similar response on the tested 8$\times$8\,cm$^{2}$ region.

\subsubsection{Digital response}

The digital response is the distribution of the number of hits. It is used to calculate efficiency $\epsilon$ and hit multiplicity \textit{m} according to:

\begin{equation}\label{eqeff}
  \epsilon = 1~-~\frac{N_{0}}{N_{\rm t}}
\end{equation}

\begin{equation}\label{eqmult}
  m = \sum_{\rm i=1}^{\rm 25}~i~\frac{N_{\rm i}}{N_{\rm t} - N_{0}}
\end{equation}

\noindent
where \textit{N}$_{\rm i}$ is the number of events with ``i'' hits and \textit{N}$_{\rm t}$ the total number of events. The sum in Eq.\,\ref{eqmult} is performed from 1 to 25 because the multiplicity is measured over regions of 5$\times$5 pads (this size is justified in section\,\ref{searchregion}). The \textit{N}$_{\rm i}$ are governed by a Binomial distribution so the errors on $\epsilon$ and \textit{m} are calculated as:

\begin{equation}\label{eqefferr}
  \sigma_{\epsilon}^{2} = \frac{N_{0}}{N_{\rm t}}~(1-\frac{N_{0}}{N_{\rm t}})
\end{equation}

\begin{equation}\label{eqmulterr}
  \sigma_{m}^{2} = \sum_{\rm i=1}^{\rm 25}~i^{2}~\frac{N_{\rm i}}{N_{\rm t}^{2}}~(1~-~\frac{N_{\rm i}}{N_{\rm t}})
\end{equation}

\noindent
When testing a given prototype, the other prototypes are used as a telescope and their mesh voltage is thus kept at 370\,V (at which an efficiency of 95\,\% is achieved, as will be shown later). In the prototype under test, the voltage was varied between 300--390\,V. At each voltage, a sample of about 2$\times$10$^{4}$ muons was recorded. The first threshold is set at roughly 6 times the noise level ($\sim$\,1.5\,fC). Second and third thresholds are set at 14\,fC and 63\,fC respectively. At the maximum tested voltage of 390\,V, these thresholds correspond to roughly 1 and 4 times the most probable value of the Landau signal distribution (\textit{i.e.} 1 and 4 MIPs). These settings are good enough to spot possible chamber-to-chamber signal non-uniformity.

\begin{figure}[htbp]
\centering
\includegraphics[width=0.85\columnwidth]{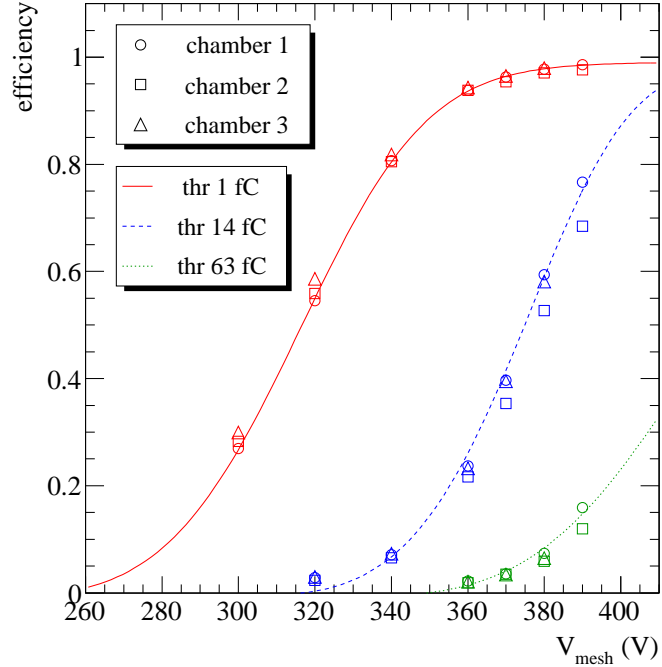}
\caption{Efficiency versus mesh voltage in 3 Micromegas chambers at 3 values of threshold. Lines are the result of a fit and are a guide to the eye.}
\label{mipresponse1}
\end{figure}

The efficiency is plotted versus voltage in Fig.\,\ref{mipresponse1}. Statistical and systematic error bars are of the size of the markers. Two observations are made. First, an efficiency of 95\,\% is achieved for the 3 prototypes at a relatively low gas gain of 2$\times$10$^{3}$ (370\,V). This was expected from the low readout threshold of $\sim$\,1.5\,fC. Secondly, at given threshold and voltage, a similar efficiency is measured in the 3 prototypes.

\begin{figure}[htbp]
\centering
\includegraphics[width=0.85\columnwidth]{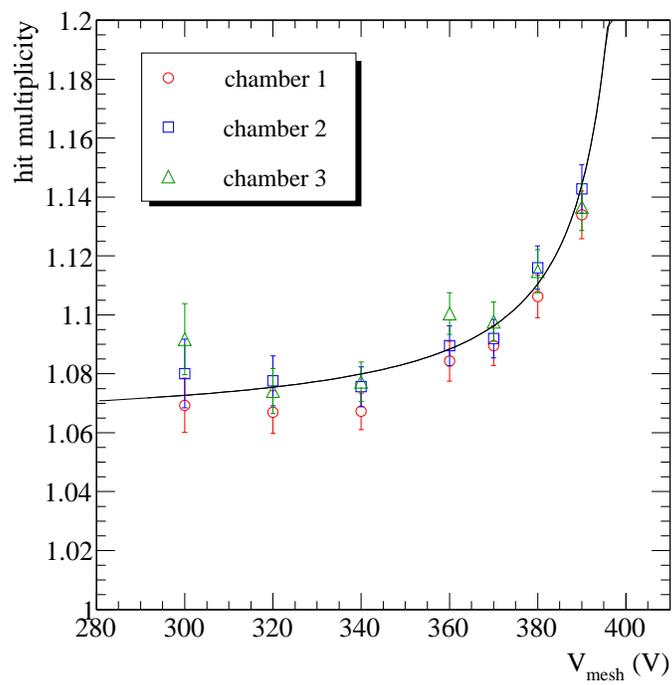}
\caption{Hit multiplicity versus mesh voltage in 3 Micromegas chambers. The fitted line is a guide to the eye.}
\label{mipresponse2}
\end{figure}

At a given voltage and for the lowest threshold, the hit multiplicity does not vary significantly in the 3 prototypes, remaining below 1.1 up to 370\,V (Fig.\,\ref{mipresponse2}). The observed variations with voltage can be explained by an increased single electron sensitivity: at higher voltages, less and less primary electrons are necessary to pass the threshold. Pads in the neighbouring of the pad crossed by a muon are thus more likely to fire and the hit multiplicity increases.

\vspace{-2mm}

\subsection{Position scan}

The scan is performed with the SDHCAL using a low-rate defocused mixed beam of muons and pions. Pions were used to study the calorimetric performance of the SDHCAL (which are not reported here) while muons were used to monitor the performance of the active layers. The data sample was collected through several runs recorded  over one day at various beam energies (20--150\,GeV). The 3 prototypes were operated at 370\,V. Event reconstruction and identification of traversing muons are detailed first, followed by track fit and response measurement.

\subsubsection{Event reconstruction}\label{evtreco}

\begin{figure}[htbp]
\centering
\includegraphics[width=0.85\columnwidth]{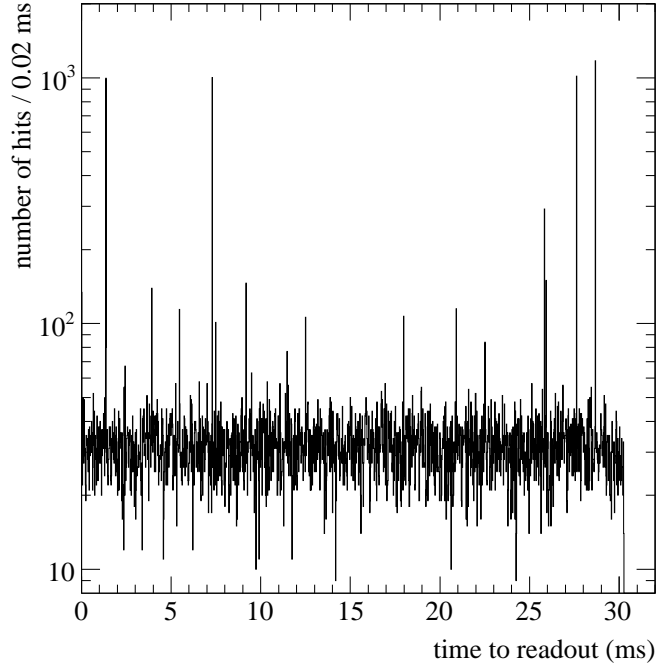}
\caption{Time spectrum of hits in the SDHCAL measured between two internally-triggered readouts.}
\label{sdhcaltimeframe}
\end{figure}

The readout of the SDHCAL is internally triggered. Between two readouts, all signals above threshold are recorded with a timestamp until one ASIC memory is full. The raw information is a collection of hits with pad and layer coordinates and a timestamp. The time spectrum of hits is composed of a uniform background corresponding to noise and sharp peaks that sign the interaction of particles inside the SDHCAL (Fig.\,\ref{sdhcaltimeframe}). The event reconstruction consists in finding the peaks and selecting the hits within.

\subsubsection{Track selection and fit}

To measure the muon response of the Micromegas prototypes, muons traversing the whole SDHCAL prototype along the beam direction are used. This requires to distinguish between three categories of events: pion showers, cosmic muons and beam muons.

The identification of showers is based on the total number of hits which, in the studied energy range, is larger for showers than for muons. Beam and cosmic muons traverse the SDHCAL horizontally and vertically respectively. They are easily distinguished using the number of hits per layer.

Above 20\,GeV, multiple scattering effects are small and the beam muon tracks are straight. Their parameters are extracted from a fit using two linear functions defined in the vertical and horizontal planes. The Micromegas hits are not included in the fits. The contamination of events by noise is negligible (an average value of $\sim$\,0.35 hit/200\,ns clock cycle is quoted in \cite{HAD13} for RPCs only). Requiring a $\chi^{2}$\,/\,\textit{NDF} value below 2 (where \textit{NDF} is the number of degree of freedom of the fit), the fit quality is good for 99.5\% of events, namely half a million events.

\subsubsection{Analysis and results}\label{searchregion}

\begin{figure}[htbp]
\centering
\includegraphics[width=0.85\columnwidth]{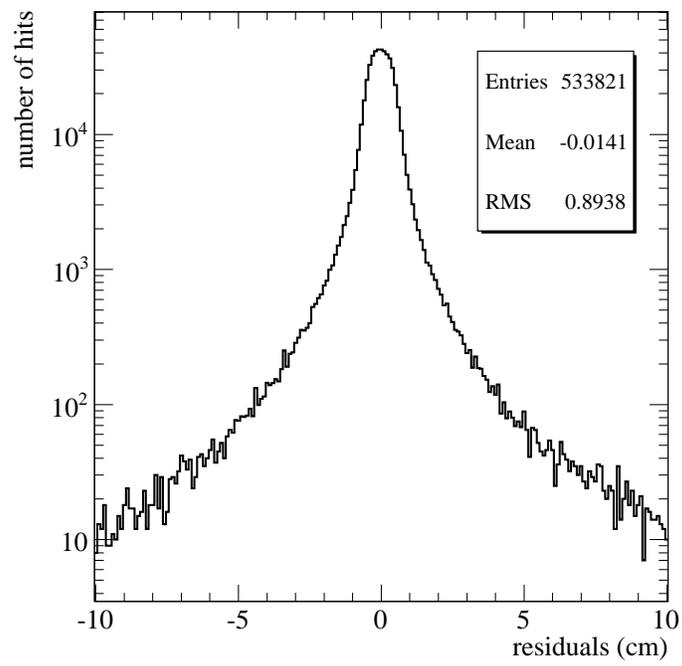}
\caption{Track fit residuals in the vertical plane of a Micromegas chamber.}
\label{trackfit2}
\end{figure}

Muon tracks are extrapolated to the Micromegas prototypes. The intercept is required to be in a fiducial area, excluding non-instrumented regions between ASUs and the 2 outermost pad rows and columns of each ASU. Then, hits are counted in a region of $\pm$2 pads around around the targeted pad. The region size is determined from the track fit residuals. Their distribution is shown in Fig.\,\ref{trackfit2}: more than 95\,\% of hits are in a region of $\pm$2 around the targeted pad. This justifies the size of the search region and also the number of excluded pad rows and columns. The digital response is measured for all ASICs and for the 3 readout thresholds. Efficiency and hit multiplicity are then calculated using Eq.\,\ref{eqeff}--\ref{eqmulterr}.

\begin{figure}[htbp]
\centering
\includegraphics[width=0.85\columnwidth]{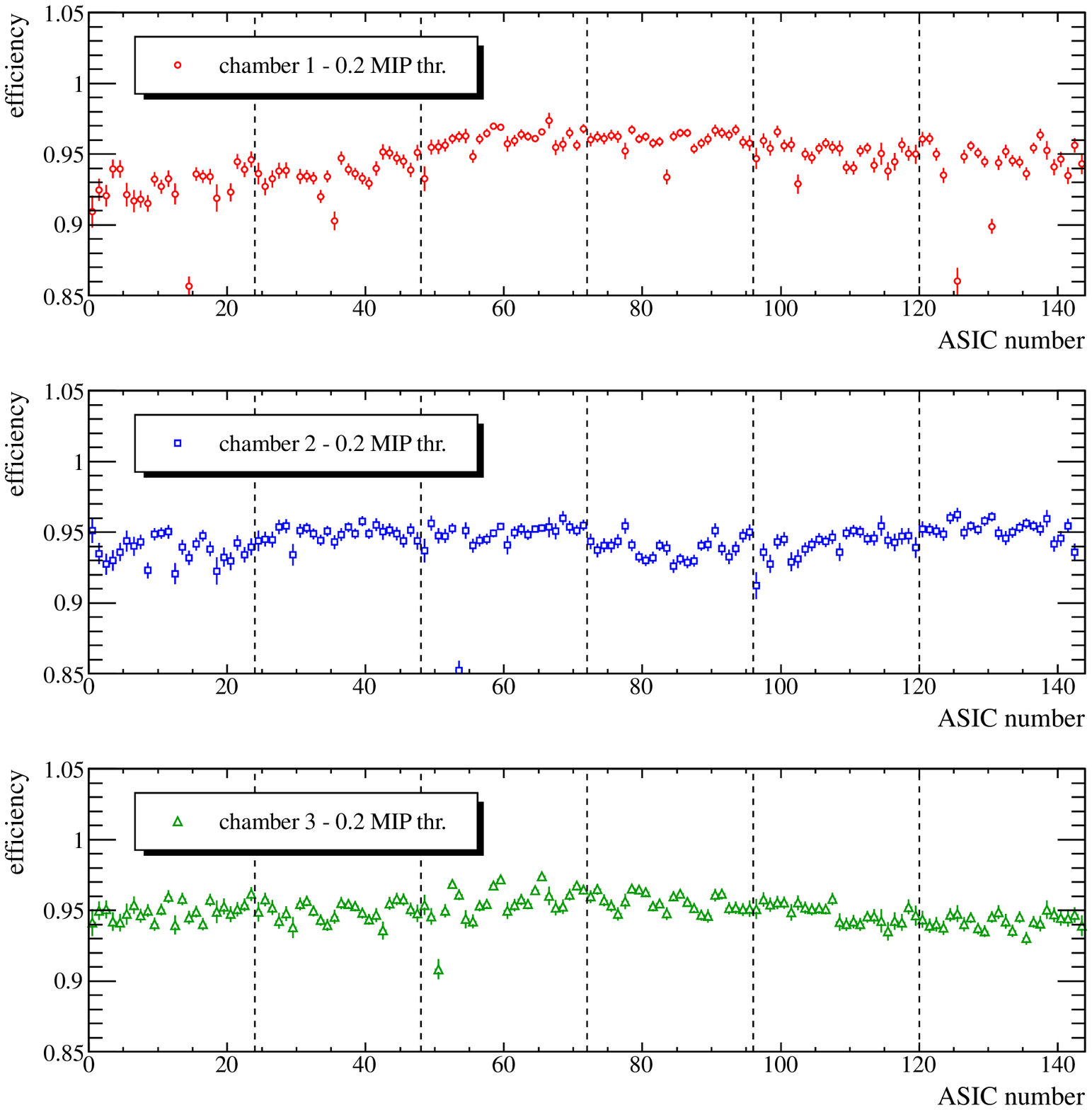}
\includegraphics[width=0.85\columnwidth]{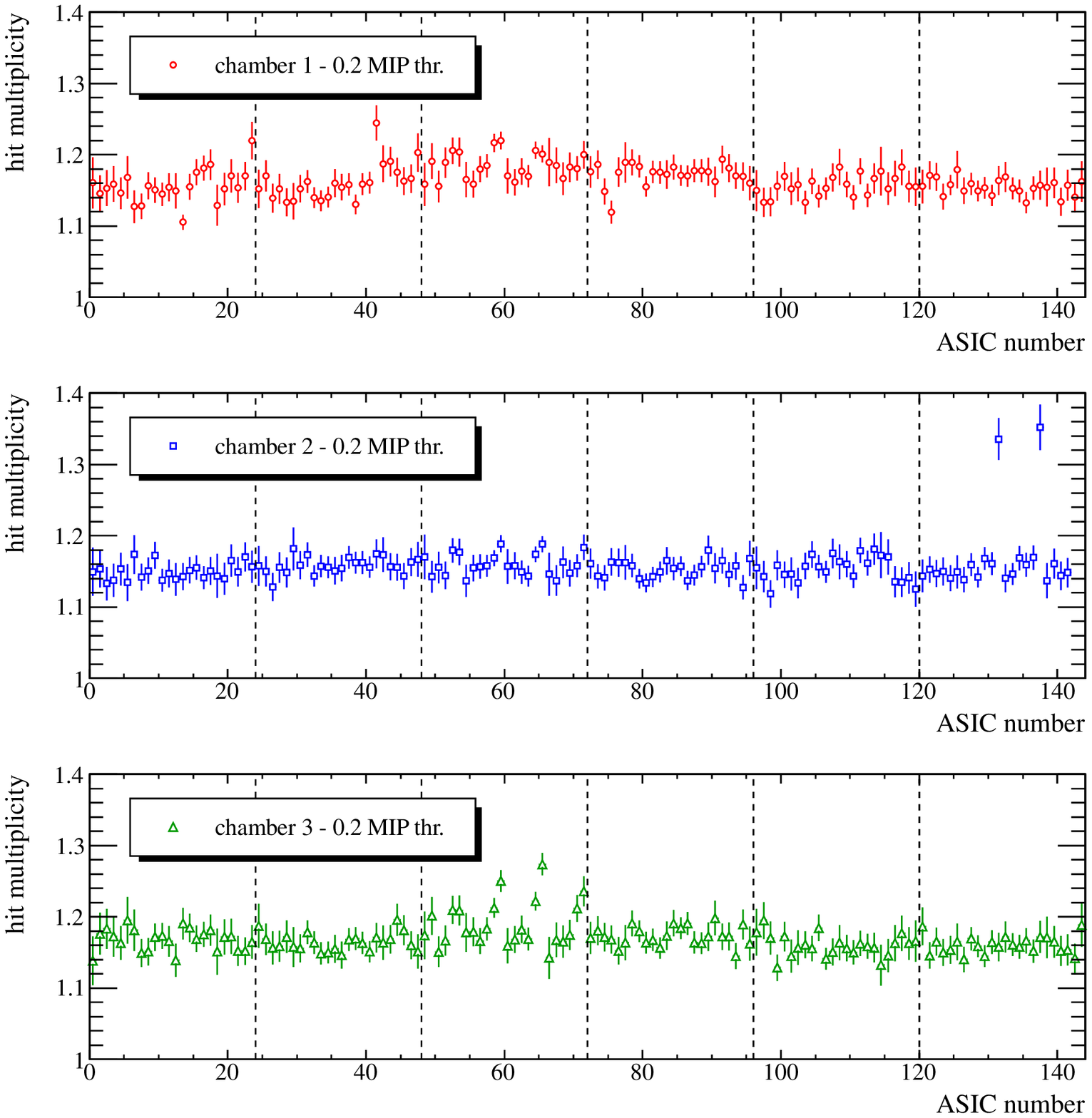}
\caption{Efficiency (3 top plots) and hit multiplicity (3 bottom plots) per ASIC for 3 Micromegas prototypes at a threshold of 0.2 MIP.}
\label{efficiency1}
\end{figure}

\begin{figure}[htbp]
\centering
\includegraphics[width=0.85\columnwidth]{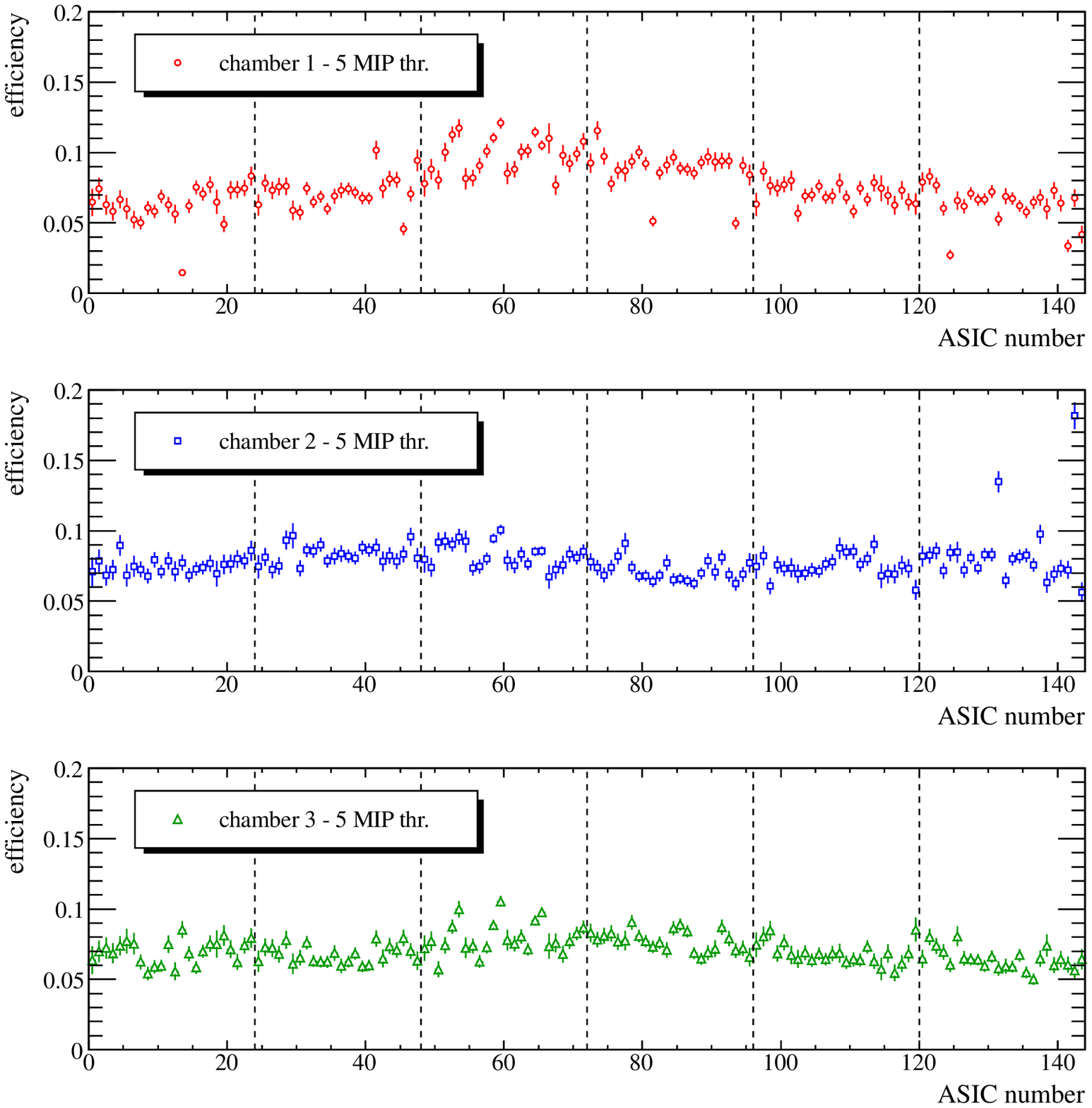}
\includegraphics[width=0.85\columnwidth]{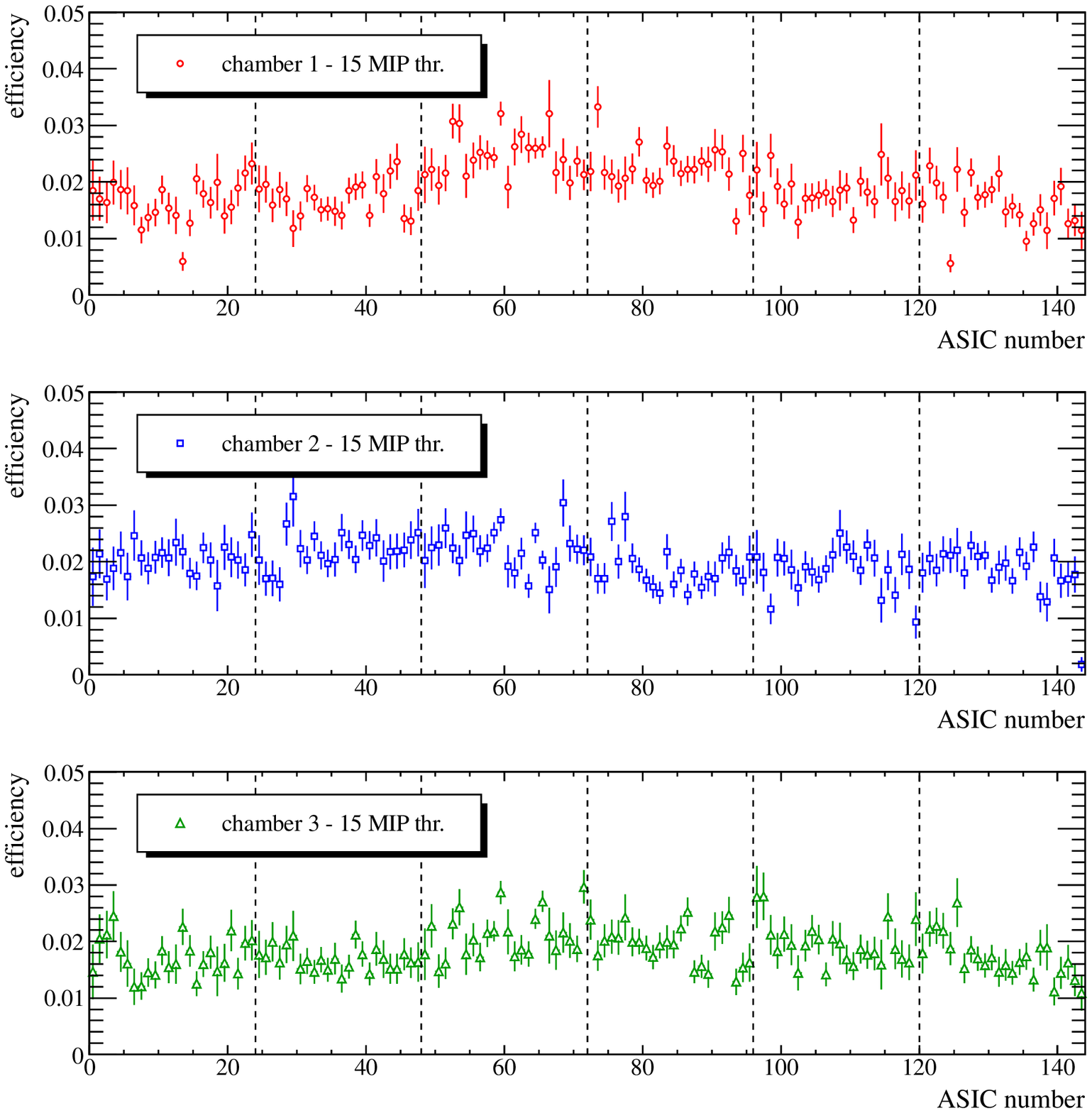}
\caption{Efficiency per ASIC for 3 Micromegas prototypes at a threshold of 5 MIP (3 top plots) and 15 MIP (3 bottom plots).}
\label{efficiency2}
\end{figure}

ASIC-to-ASIC variations of efficiency and hit multiplicity are shown in Fig.\,\ref{efficiency1} and \ref{efficiency2}. Data from different ASUs are separated by vertical dashed lines. For a given ASU (or Bulk mesh), the ASIC number runs along the vertical axis (4 columns of 6 ASICs). Average values and dispersions deduced from a gaussian fit are listed in Table\,\ref{efftab}. At low threshold (0.2\,MIP), efficiency and hit multiplicity are close to 95\,\% and 1.15 respectively, with variations of 1--2\,\% for all prototypes. At higher thresholds (5 and 15 MIPs), one is more sensitive to possible detector non-uniformity but efficiency variations remain small. Such good uniformity implies that the calibration of a Micromegas sampling calorimeter should be simple and its constant term small.

Although dispersions are small, the measured trends exhibit a periodic pattern, most pronounced for the 2 central ASUs. The efficiency increases slightly towards the centre of the prototype. A likely explanation is an increased temperature (and thus increased gas gain) at the centre of the module.

\begin{table}[htb]
\renewcommand{\arraystretch}{1.25} 
\begin{center}
\begin{tabular}{c|cccccc}
  & chamber 1 	 & chamber 2 	 & chamber 3 \\
\hline
$\epsilon_{\rm~0.2 MIP}$ (\%) & 95.0\,$\pm$\,1.6 & 94.4\,$\pm$\,0.9 & 94.9\,$\pm$\,0.8 \\
$\epsilon_{\rm~5 MIP}$ (\%) & 7.3\,$\pm$\,1.8 & 7.7\,$\pm$\,0.8 & 7.0\,$\pm$\,1.1 \\
$\epsilon_{\rm~15 MIP}$ (\%) & 1.9\,$\pm$\,0.5 & 2.1\,$\pm$\,0.4 & 1.8\,$\pm$\,0.3 \\
$\textit{m}_{\rm~0.2 MIP}$ & 1.16\,$\pm$\,0.02 & 1.15\,$\pm$\,0.02 & 1.17\,$\pm$\,0.02 \\
\end{tabular}\\[2pt]
\caption{Efficiency and hit multiplicity of 3 Micromegas chambers (mean\,\,$\pm$\,\,RMS).}
\label{efftab}
\end{center}
\end{table}

\vspace{-7mm}

\section{Noise study}

In a digital calorimeter, noise will degrade the energy resolution if the resulting number of hits is not negligible compared to the number of hits from showering particles. Moreover, the detectors at a future linear collider will operate in a self-trigger mode. This imposes that noise should not saturate the memory of the front-end ASICs during the bunch trains. The relevant quantity is therefore the noise rate per ASIC. It was measured with the SDHCAL so the experimental conditions are as realistic as possible. Moreover, the SDHCAL particle identification capability can be used to separate hits from cosmic particles and hits from noise. This is important because, as will be shown, noise rates in the Micromegas prototypes are small.

\subsection{Analysis}

As detailed in section\,\ref{evtreco}, events occurring in the SDHCAL are identified as sharp peaks in the time spectrum of hits (Fig.\,\ref{sdhcaltimeframe}). It is assumed that hits recorded between two events originates purely from noise. This hypothesis is probably good enough as RPCs and Micromegas in the SDHCAL are almost transparent to neutrons and insensitive to possible delayed-neutron backgrounds. Noise rates are measured by summing the number of hits and the time between events and taking their ratio. When defining the gaps between events, a time window of $\pm$\,2\,$\mu$s around the event peaks is applied.

\begin{figure}[htbp!]
\centering
\includegraphics[width=0.85\columnwidth]{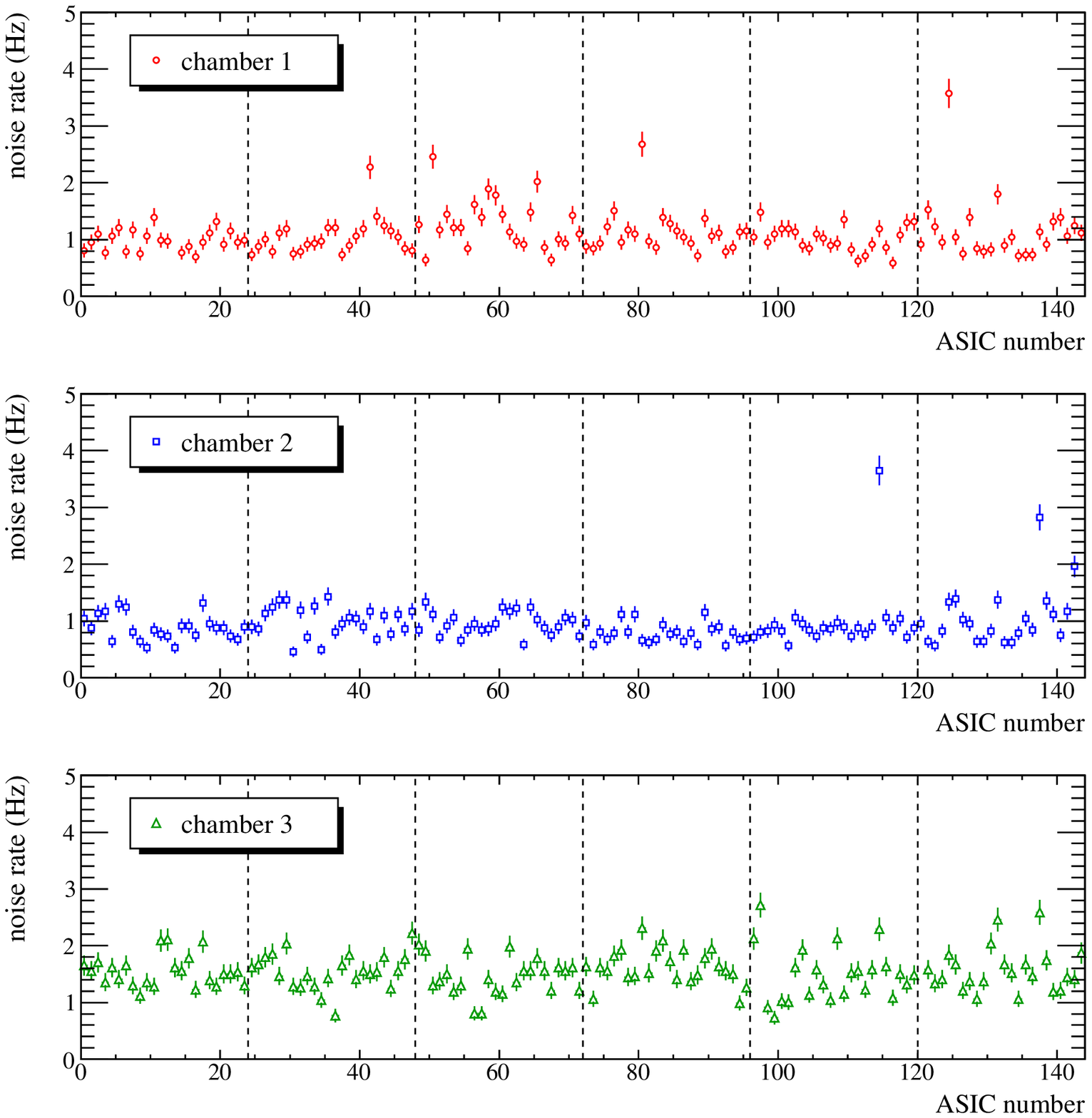}
\includegraphics[width=0.85\columnwidth]{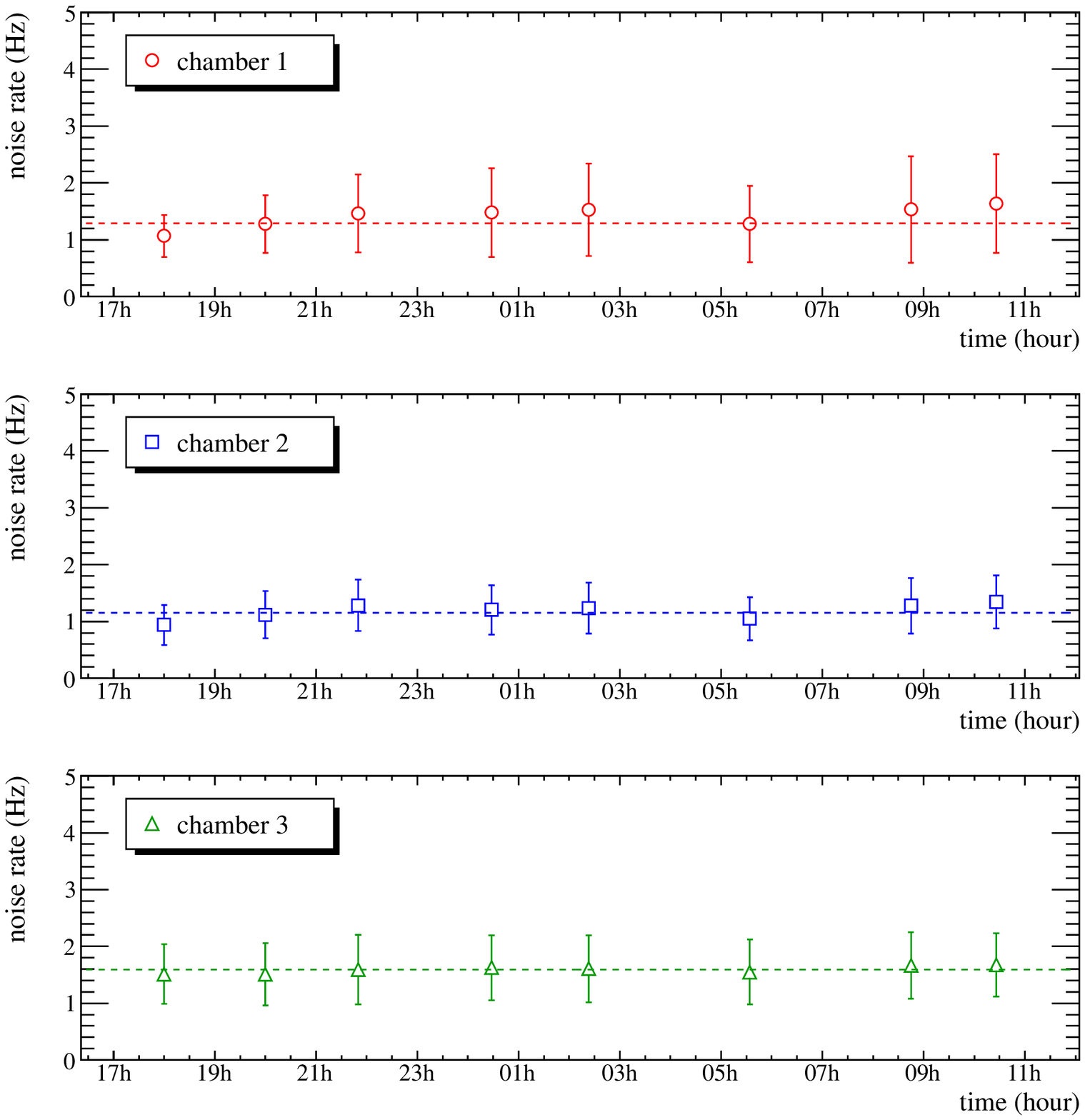}
\caption{Noise rate per ASIC of the 3 prototypes during a one hour long run (3 top plots). Average rate as a function of time for several runs (3 bottom plots), the errors bars indicate the RMS dispersion.}
\label{noiserate}
\end{figure}

\subsection{Results}

The noise rate was measured for each ASIC of the 3 prototypes using the same data sample previously used for the position scan (mixed muon/pion beam). This sample contains data from 8 consecutive runs recorded at various energies (20--150\,GeV). The results of a run appear in Fig.\,\ref{noiserate} (top). The rate is uniform within a prototype and from one prototype to the other. The stability with time is excellent as seen in Fig.\,\ref{noiserate} (bottom) where the average rate per run is plotted (averaging is done over all ASICs). An average noise rate of 1--2\,Hz/ASIC is found.

An ASIC memory can record up to 127 events. The duration of bunch trains at the ILC (1\,ms) is then negligible compared to the time to fill one memory entirely with noise (127\,s at 1\,Hz/ASIC). In addition, the noise rate per prototype (144 ASIC) is roughly 300\,Hz. Considering a shower developing in a Micromegas SDHCAL of 50 layers, the number of noise hits expected in the typical time window of an event (1\,$\mu$s) would thus be less than one ($\sim$\,10$^{-2}$ without fiducial cuts). In conclusion, the noise level in the Micromegas prototypes complies with ILC specification and would yield a zero noise term in a Micromegas digital calorimeter.

\section{Stability study in high-energy pion showers}

The stability of the prototypes is assessed by measuring their rate capability and spark probability in high-energy pion showers. This is done by measuring the digital response and monitoring the mesh currents at different shower rates. This study is best conducted with the standalone setup and a beam of 150\,GeV pions collimated to a spot of $\sim$\,1\,cm$^{2}$. With this setup, most pions shower inside a 2\,$\lambda_{\rm int}$ thick iron absorber. A fraction of them, however, traverses the absorber without showering in which case a single track is measured in the prototypes downstream. The identification of penetrating pions is necessary to estimate the shower rate and is detailed first. Then, the response is measured at different mesh voltages to determine the voltage at which performing the rate scan. Finally, results from the rate scan are presented.

\subsection{Digital response}

\begin{figure}[htbp]
\centering
\includegraphics[width=0.8\columnwidth]{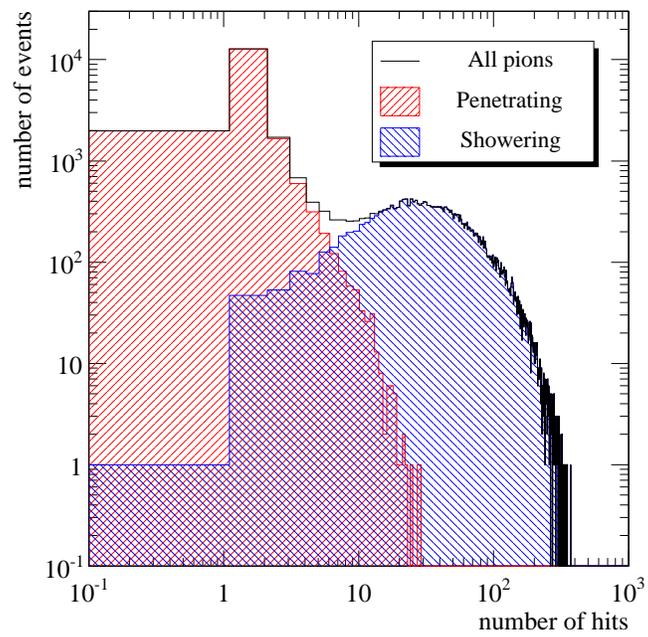}
\caption{Number of hits from penetrating and showering pions in the first Micromegas chamber downstream the absorber. The first bin lower edge in fact equals zero.}
\label{figure_nhit_shower1}
\end{figure}

At a mesh voltage of 370\,V in all prototypes, the distribution of the number of hits per triggering pion was measured. The distribution measured in the first prototype downstream the absorber is shown in Fig.\,\ref{figure_nhit_shower1}. It exhibits a sharp peak at 1 hit and a long tail from penetrating and showering pions respectively. The contribution from showering pions is extracted by requiring more than 3 hits in the other prototypes. It appears in Fig.\,\ref{figure_nhit_shower1} together with the remaining contribution from penetrating pions. According to this selection cut, the fraction of pions that shower before traversing the first prototype is $\sim$\,90\,\%, as expected from the absorber thickness of 2\,$\lambda_{\rm int}$. This fraction is later used to deduce the shower rate from the beam rate.

\subsection{Voltage scan}

\begin{figure}[htbp]
\centering
\includegraphics[width=0.8\columnwidth]{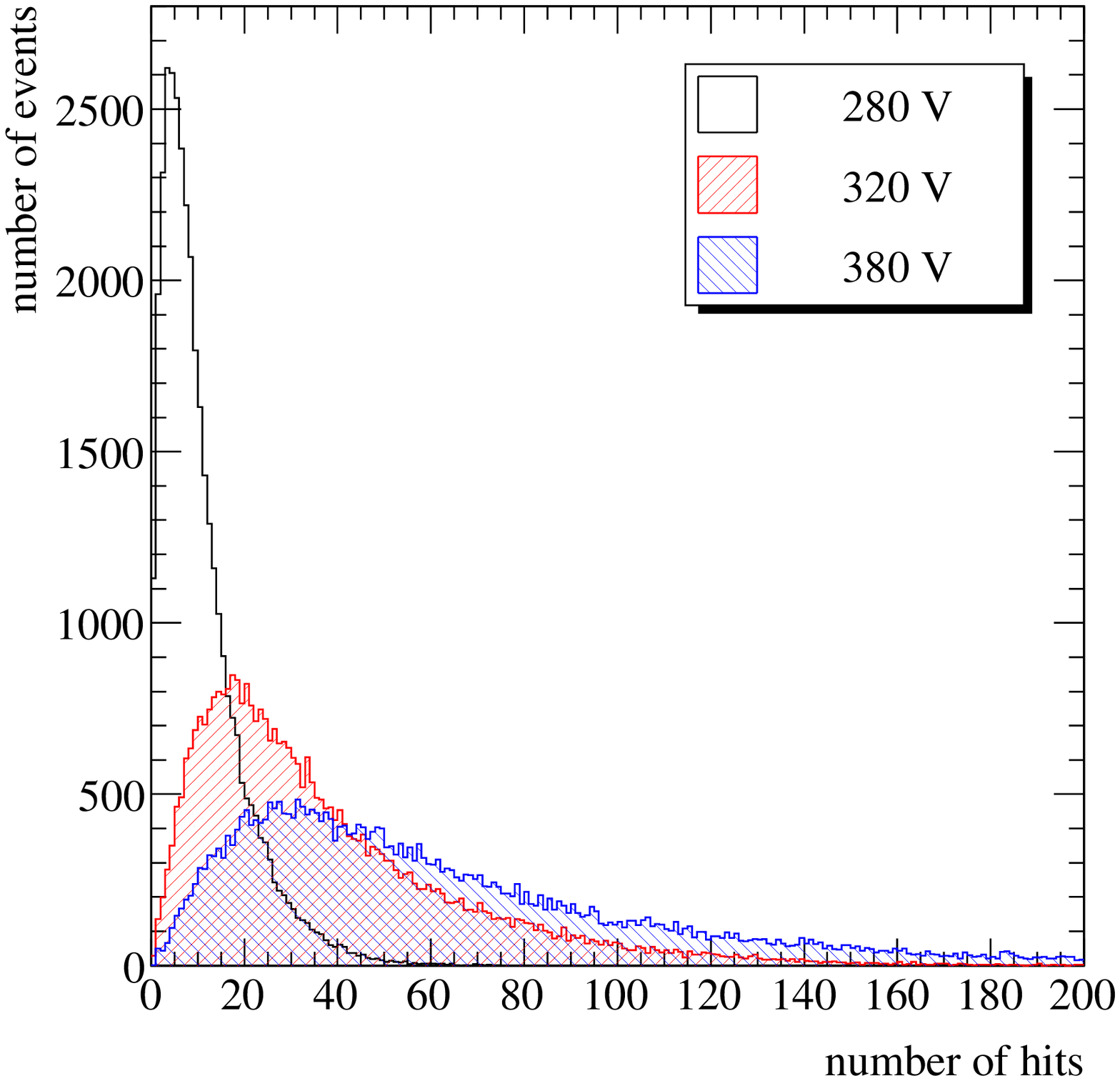}
\caption{Number of hits from showering pions at various mesh voltages in the first Micromegas chamber downstream the absorber.}
\label{figure_nhit_shower2}
\end{figure}

About 5$\times$10$^{4}$ pion triggers were recorded at different values of the mesh voltage in the first prototype (280--380\,V). In the other prototypes, the voltage was kept at 370\,V in order to identify penetrating pions. The distribution of the number of hits from showering pions measured in the first prototype is shown in Fig.\,\ref{figure_nhit_shower2}.

\begin{figure}[htbp]
\centering
\includegraphics[width=0.76\columnwidth]{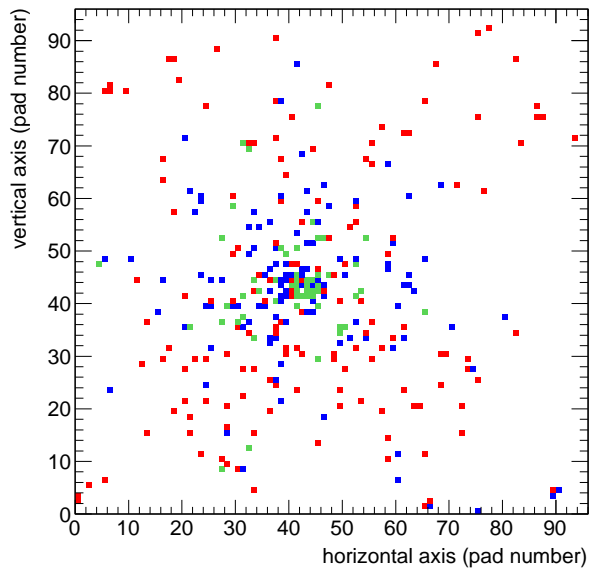}
\includegraphics[width=0.76\columnwidth]{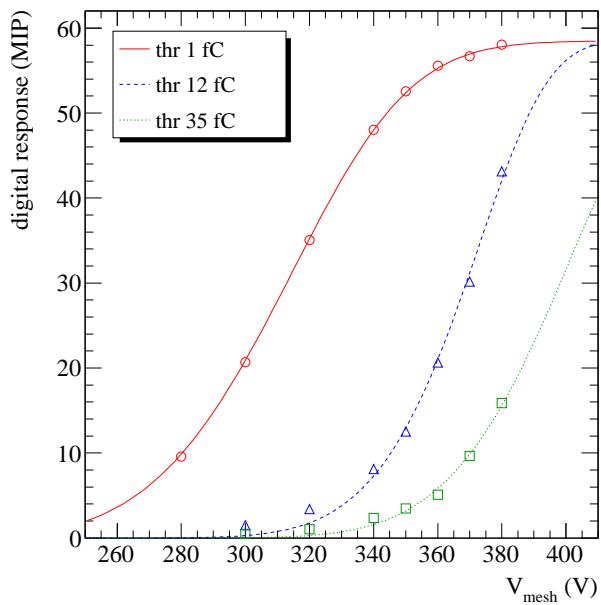}
\caption{High-multiplicity pion shower event (top). Ratio of the number of hits from showering pions and high-energy muons versus mesh voltage (bottom). Lines are the result of a fit and are a guide to the eye.}
\label{figure_event_display}
\end{figure}

 At low mesh voltage (\textit{e.g.} 280\,V), only the dense and narrow electromagnetic core is observed and a few hits are recorded. As the voltage is increased, single tracks forming the hadronic halo are progressively seen and the distribution tail expands towards higher values. The picture of a dense core surrounded by a halo of tracks is illustrated in Fig.\,\ref{figure_event_display} (top) with a display of a high-multiplicity event.

The mean value of the distribution keeps increasing up to the highest tested voltage (380\,V). This is likely due to the increase of the single-particle hit multiplicity (Fig.\,\ref{mipresponse2}) which is therefore used as normalisation. The ratio expressed in MIP units in fact reaches a plateau. As shown in Fig.\,\ref{figure_event_display} (bottom) an asymptotic value of 58\,MIPs is suggested, meaning that 95\,\% of the shower signal is seen at 360\,V. Measurements at higher thresholds are also included in the plot. By comparing the low-threshold efficiency curves in Fig.\,\ref{mipresponse1} and \ref{figure_event_display} (bottom), a similar shape is observed: a given fraction of the asymptotic value is reached at roughly the same voltage (\textit{e.g.} 50\,\% at 310--320\,V). This suggests that at this energy the hadron shower signal (a sum of hits) is mainly caused by single particles rather than by the narrow electromagnetic core.

\subsection{Rate scan}

\subsubsection{Rate capability}

The scan was performed at a voltage of 370\,V in all prototypes so the shower response is close to its maximum. The pion rate is measured with two scintillators attached to the absorber and was varied from 1 to 30\,kHz (the beam spot is $\sim$\,1\,cm$^{2}$). The shower rate is then obtained from the known fraction of showering pions ($\sim$\,90\,\%). For each run at different rate, the number of hits per event was measured. Its mean value is determined for the 3 prototypes at 3 threshold values. To clearly observe a possible drop of efficiency with rate, the mean value at given rate and threshold and in a given prototype is normalised to the mean value at lowest rate (1\,kHz) and threshold ($\sim$\,1.5\,fC) in this prototype.

\begin{figure}[htbp]
\centering
\includegraphics[width=0.8\columnwidth]{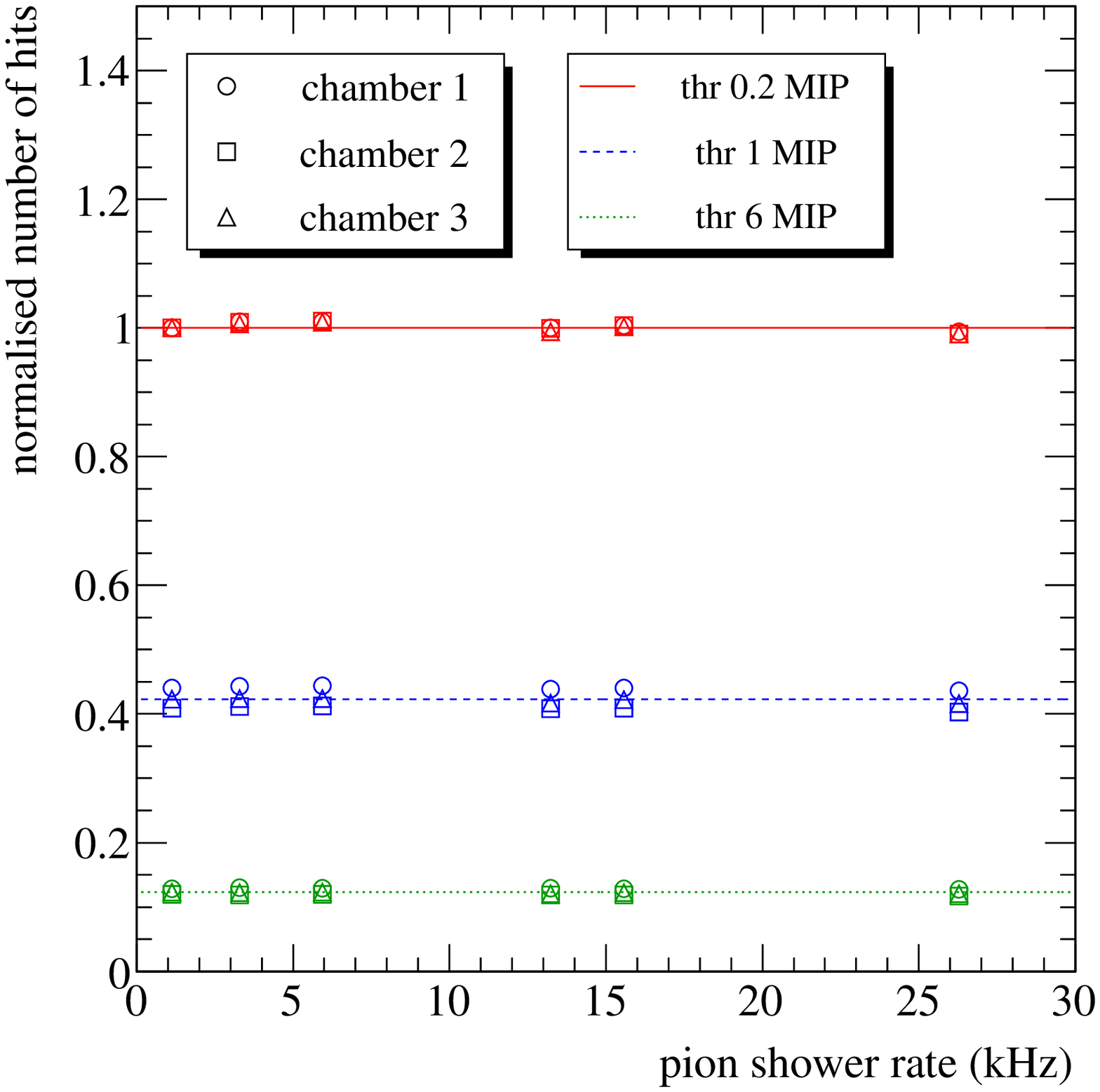}
\caption{Micromegas chamber response versus pion shower rate after normalisation to the response at 1\,kHz and 0.2\,MIP threshold. The horizontal lines are a fit of a constant to the data points.}
\label{figure_normnhit_rate}
\end{figure}

The response is plotted versus shower rate in Fig.\,\ref{figure_normnhit_rate}: variations are negligible up to the highest shower rate of 26\,kHz and for the 3 thresholds. This last point is important because for the lowest threshold value, a possible rate-induced loss of gas gain may be too small to significantly impact on the efficiency. On the other hand, at higher thresholds (\textit{e.g.} 1\,MIP), the sensitivity of the efficiency to gas gain variations is enhanced. The fact that the efficiency at higher threshold is constant implies that the prototypes are free of space-charge effects over the tested rate range.

As previously measured, the digital reponse to showering pion at 370\,V is close to 58 MIPs. Accordingly, at the maximum rate of 26\,kHz, at least 1.5$\times$10$^{6}$ particles are traversing or stopping in the first prototype every second. Averaging over the whole area, a rate per unit area of $\sim$\,150\,Hz/cm$^{2}$ is obtained. This value is obviously misleading because it does not account for the transverse position distribution of the shower secondary particles around the shower axis (which peaks towards the axis) and for the fact that several particles can cross a pad and still be counted as one hit. Calculation of the effective rate would probably require inputs from Monte Carlo simulation and is beyond the scope of this study.

\subsubsection{Spark probability}

\begin{figure}[htbp]
\centering
\includegraphics[width=0.8\columnwidth]{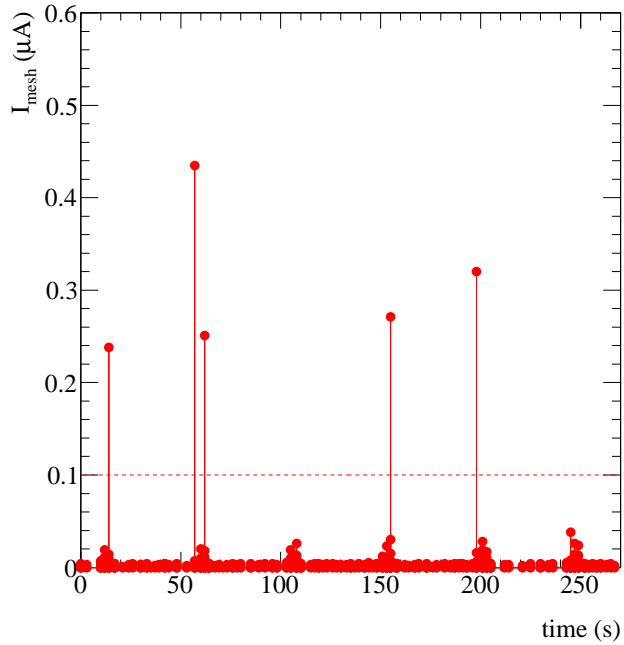}
\caption{Mesh currents of a Micromegas chamber recorded at a pion shower rate of 26\,kHz. The horizontal line indicates the applied threshold to count sparks.}
\label{sparkproba1}
\end{figure}

Current and voltage variations of the prototype meshes and drift electrodes were monitored during the tests with an accuracy of $\pm$1\,nA and $\pm$0.1\,V. The used slow-control system actually records any variation of current or voltage larger than the given accuracies. When the beam is turned off and at a mesh voltage of 370\,V, the current drawn through the power-supply is generally of the order of a few nA. When the beam is turned on, the current increases up to $\sim$\,25\,nA at the maximum rate. A spark occuring between a mesh and a pad will produce a sudden spike of current, typically a few hundred nA. This is illustrated in Fig.\,\ref{sparkproba1} which shows the current history of a prototype (6 meshes) at the highest rate of the scan. The SPS spill structure is seen as well as spikes during spills.

\begin{figure}[htbp]
\centering
\includegraphics[width=0.95\columnwidth]{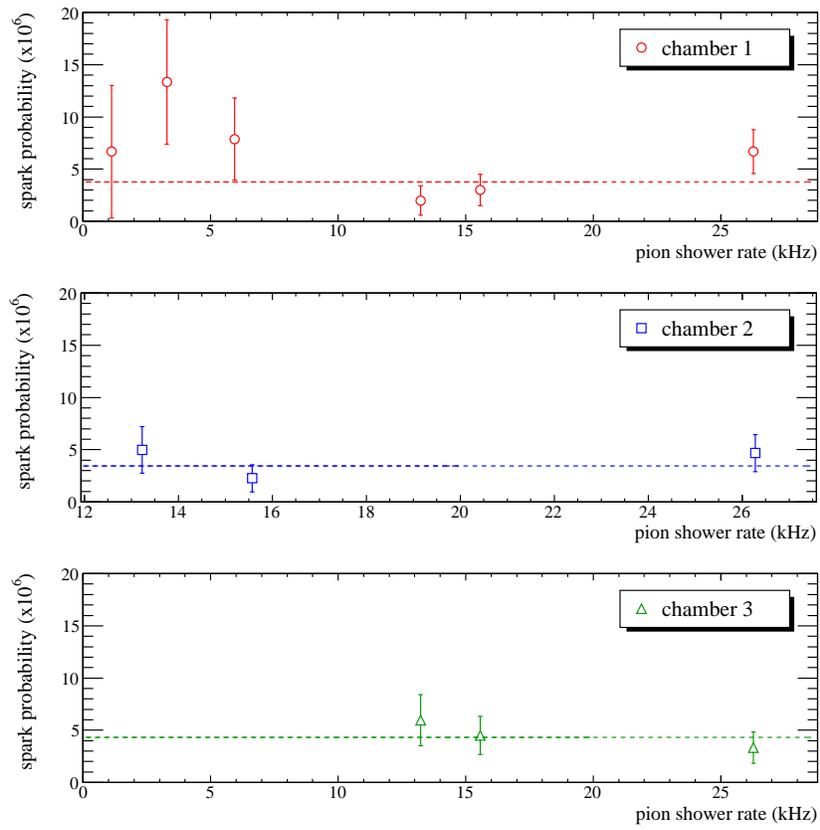}
\caption{Spark probability per showering pions as a function of shower rate for 3 Micromegas chambers. The horizontal lines are a fit of a constant to the data points.}
\label{sparkproba2}
\end{figure}

Sparks are identified as a current passing a certain threshold. The spark probability per showering pions is then calculated as the number of sparks divided by the expected number of showering pions. The latter is estimated from the number of triggering pions during a run and the fraction of showering pions, namely 90\,\%. The current threshold should be higher than the typical current activity during the spills. The threshold was varied between 25\,nA and 100\,nA with little effects on the measured probability. The probability per showering pion is shown in Fig.\,\ref{sparkproba2} for a threshold of 100\,nA and is independent of rate, as expected. The average probability in the 3 prototypes is low: $\sim$\,5$\times$10$^{-6}$ per showering pions.

In a sampling calorimeter, the spark probability should on average be highest at the shower maximum. It should therefore depend on the layer number. With the standalone setup, only 4\,mm of steel (chamber cover and baseplate) and 10\,cm air separate the prototypes. This is probably too little to significantly reduce the particle flux from one prototype to the next. Therefore the spark probability is similar in the 3 prototypes. If a constant probability of 5$\times$10$^{-6}$ per shower is assumed for all layers, an upper limit on the spark rate can be derived. In a 50 layers calorimeter, it is equal to 2.5$\times$10$^{-4}$ which is still fairly small. At a future linear collider experiement, however, showers will come into jets and physics events of interest will contain several jets. In these conditions, it is unclear if sparking will be problematic. Nevertheless, new detector designs incorporating resistive layers are under study by the Micromegas community to avoid sparking \cite{ALE11}. These designs will be adapated for calorimetry in the future.

\section{Conclusion}

Three advanced large-area Micromegas prototypes were constructed for a semi-digital hadron calorimeter (SDHCAL). During a testbeam campaign at CERN, several possible issues concerning Micromegas calorimetery were addressed. The achieved performance are very satisfying and probably sufficient for a linear collider SDHCAL: low noise, high efficiency, low hit multiplicity, uniform response, high rate capability and small discharge probability. Based on these findings, the energy resolution of a Micromegas calorimeter is expected to be close to the stochastic limit given by sampling and intrinsic fluctuations.

\section*{Acknowledgements}

The authors wish to thank the french \textit{Agence Nationale de la Recherche} (ANR) for financial support, the SDHCAL group of the CALICE collaboration for its help in manning the shifts during the testbeam and the RD51 CERN laboratory for providing useful hardware.

\end{document}